\documentclass[preprint,superscriptaddress,floatfix,prl]{revtex4}
\usepackage{graphicx}
\usepackage{textcomp}
\usepackage{units}
\usepackage{siunitx}
\usepackage{amsmath}
\usepackage{amsfonts}
\usepackage{amssymb}

\usepackage[usenames,dvipsnames,svgnames,table]{xcolor}
\usepackage[normalem]{ulem}

\renewcommand{\vec}[1]{\boldsymbol{#1}}

\begin{document}

%\preprint{APS/123-QED}

\title{Orbitally resolved superconductivity in real space: FeSe}


\author{Fang Yang}
\email{francoyang1988@163.com}
\affiliation{Institute for Nanoelectronic Devices and Quantum Computing, Fudan University, Songhu Rd. 2005, 200438 Shanghai, P.R. China}
\affiliation{Physikalisches Institut, Karlsruhe Institute of Technology, Wolfgang-Gaede Str. 1, 76131 Karlsruhe, Germany}

\author{Jasmin Jandke}
\affiliation{Physikalisches Institut, Karlsruhe Institute of Technology, Wolfgang-Gaede Str. 1, 76131 Karlsruhe, Germany}

\author{Peter Adelmann}
\affiliation{Institut f\"ur Festk\"orperphysik, Karlsruhe Institute of Technology, 76344 Karlsruhe, Germany}

\author{Markus J. Klug}
\affiliation{Institut f\"ur Theorie der Kondensierten Materie, Karlsruhe Institute of Technology, Wolfgang-Gaede Str. 1, 76131 Karlsruhe, Germany}

\author{Thomas Wolf}
\affiliation{Institut f\"ur Festk\"orperphysik, Karlsruhe Institute of Technology, 76344 Karlsruhe, Germany}

\author{Sergey Faleev}
\affiliation{Max-Planck-Institut f\"{u}r Mikrostrukturphysik, Weinberg 2, D-06120 Halle, Germany}

\author{J\"{o}rg Schmalian}
\affiliation{Institut f\"ur Theorie der Kondensierten Materie, Karlsruhe Institute of Technology, Wolfgang-Gaede Str. 1, 76131 Karlsruhe, Germany}
\affiliation{Institut f\"ur Festk\"orperphysik, Karlsruhe Institute of Technology,
76344 Karlsruhe, Germany}
\author{Matthieu Le Tacon}
\affiliation{Institut f\"ur Festk\"orperphysik, Karlsruhe Institute of Technology,
76344 Karlsruhe, Germany}

\author{Arthur Ernst}
\affiliation{Max-Planck-Institut f\"{u}r Mikrostrukturphysik, Weinberg 2, D-06120 Halle, Germany}
\affiliation{Institute for Theoretical Physics, Johannes Keppler University Linz, Altenberger Stra{\ss}e 69, 4040 Linz, Austria}

\author{Wulf Wulfhekel}
\affiliation{Physikalisches Institut, Karlsruhe Institute of Technology, Wolfgang-Gaede Str. 1, 76131 Karlsruhe, Germany}
%\affiliation{State Key Laboratory of Surface Physics and Department of Physics, Fudan University, Shanghai 200433, China}

\date{\today}

\begin{abstract} {\bf Multi-orbital superconductors combine
    unconventional pairing with complex band structures, where
    different orbitals in the bands contribute to a multitude of
    superconducting gaps.  We here demonstrate a fresh approach using
    low-temperature scanning tunneling microscopy (LT-STM) to resolve
    the contributions of different orbitals to superconductivity. This
    approach is based on STM's capability to resolve the local density
    of states (LDOS) with a combined high energy and sub unit-cell
    resolution. This technique directly determines the orbitals on
    defect free crystals without the need for scatters on the surface
    and sophisticated quasi-particle interference (QPI) measurements.
    Taking bulk FeSe as an example, we directly resolve the
    superconducting gaps within the units cell using a 30\ mK STM.  In
    combination with density functional theory calculations, we
    are able to identify the orbital character of each gap.}

\end{abstract}

\maketitle
In this Letter, we show that by reducing the distance of the tip to the surface,
orbitally resolved information on the gap can be obtained by lateral
variations of the tunneling spectra within the unit cell. This
technique may be very helpful to gain additional information on many
multi-band superconductors and the direct observation of the orbitals
may clarify some controversies regarding the nature of the involved
bands.
% To analyze the tunneling conductance maps, We developed a method to
% rule out the first order uniaxial anisotropy of tip apex state, when
% the sample surface has a two-fold rotational symmetry. Thus, we
% obtained the real space distribution of superconducting coherence
% peaks without artifacts. The experimentally measured DOS exhibit
% identical features as partial DOS regarding orbitals of
% first-principle calculations. Further, we obtained the weight of
% different orbitals by fitting the calculated DOS by the experimental
% one. The shown results might open the door for the exact
% determination of the Wannier function in the future in order to pin
% down reliable tight-binding models for the simplest iron-based
% superconductor FeSe.
As a model system to demonstrate this approach, we
chose bulk FeSe as it has the simplest crystalline structure among
iron-based superconductors and does not require doping to become
superconducting~\cite{Hsu2008,Hirschfeld2011,B_hmer_2017,Mizuguchi2010,Coldea2018}.
The unit cell of FeSe contains a layer of two Fe atoms and
two layers Se atoms, one above and one below the Fe plane and the multi-band electronic structure
near the Fermi level consists of three hole pockets at the
$\Gamma$-point and two electron pockets at the
M-point~\cite{Watson_2015,McQueen_2009,Watson2017,Massat2016,Yamakawa2016,Ishizuka2018}.
It is established that the
$d_{xz},d_{yz},d_{xy}$ orbitals of Fe dominate the Fermi surface.
However, nematic order adds complexity to this material with a structural
phase-transition at $T_s=$90~K~ from a
high-temperature tetragonal phase ($a=b\neq c$, space group: $P4/nmm$)
to a low-temperature orthorhombic phase ($a\neq b\neq c$, space group:
$Cmma$) with no magnetic ordering.
Due to the
0.3\% orthorhombic distortion, the unit cell is only of $C_2$
rotational symmetry with two inequivalent mirror planes (see Figure
\ref{Fig1}b).
As a consequence, the
degeneracy between the $d_{xz}$ and $d_{yz}$ orbitals is lifted. A splitting of 50~meV was
observed with angle-resolved photoemission spectroscopy
(ARPES)~\cite{Zhang_2015,Watson_2015,Watson_2016,Nakajama_2014,Maletz_2014,Shimojima_2014,Suzuki_2015}.
The size of the splitting is larger than expected for an orthorhombic
lattice distortion~\cite{Suzuki_2015,Watson_20152,Watson_20172,Watson_2015} and was thus attributed to electronic nematicity~\cite{Shimojima_2014,Suzuki_2015,Kostin2018}.

Ultimately, a complex nature of superconductivity in FeSe arises, that occurs below a critical temperature of $\approx$8~K~\cite{Hsu_2008}.
The superconducting gap symmetry of
single-crystalline FeSe has been discussed widely in the literature.
Some report a nodal gap \cite{Song_2011,Kasahara_2014} and more recent
papers a nodeless pairing symmetry~\cite{Jiao_2016,Sprau2017, Kreisel2017,Benfatto2018}. Even
though there are some variations with respect to the size of the
various gaps, a uniform observation is the multigap structure. Using
QPI patterns, Sprau et al. could identify two anisotropically gapped
bands without nodes of mainly $d_{yz}$ character \cite{Sprau2017}. However, no consensus about the exact
orbital contributions of each gap has been reached~\cite{Kreisel_2015,Mukherjee_2015, Benfatto2018}.
To address these questions, we focus on STM images and spectra  spatially resolved within the unit cell.

Figure \ref{Fig1}a shows an STM topography of the sample cleaved in
situ at low temperatures. The white dots reflect the upper (or lower)
Se atoms of the surface layer, depending on the tunneling conditions
\cite{Kreisel2016}. The regular pattern illustrates the translational
symmetry of the lattice of FeSe with practically no defects.  To first
order, the tunneling current in STM is given by the LDOS of the sample
leaking out into the vacuum integrated over the bias energy window, as
has been shown by Tersoff and Hamann \cite{Tersoff1985}.  Thus, STM
topography in constant current mode reflects the iso-surface of the
energy integrated LDOS. For low sample bias, the LDOS of a normal conductor
does not vary significantly and can be considered a constant. Thus,
STM topography at low bias directly represents the iso-surface of the
electron density at the Fermi level $E_F$ decaying into the vacuum. In
the superconducting state, only the quasi-particle states outside the
gap $\Delta$ contribute to the LDOS. If the bias voltage $U$ is set
significantly above $\Delta$, the integrated LDOS in the
superconducting state is nearly that of the normal state.  In that
respect, the STM image reflects the LDOS near $E_F$ of the normal state.
In theory, the LDOS of the normal state is given by the single
particle excitations of the electrons, i.e. by the single-particle spectral function weighted by the form factor of Bloch states, at
the respective energy. These states take into account the translational symmetry
of the crystal as they are a product of a lattice periodic wave function
$u_\lambda^{\vec{k}} (\vec{r})$ and a plane wave $e^{i\vec{kr}}$, where
$\vec{k}$ is the wave vector, $\vec{r}$ is the position in real space
and $\lambda$ denotes the quantum numbers related to the atomic states
as the orbital degrees of freedom or the spin.  The band structure describes the
energy $E$ as function of $\vec{k}$ and $\lambda$. In order to study
multi-band superconductors, either ARPES is used to directly measure $E(\vec{k})$.
Alternatively, laterally resolved STM spectra of the sample including
scatterers on the surface causing quasi-particle interference patterns
can be imaged. By Fourier transforming the observed QPI patters into
momentum space, information on the bands can be obtained. Since QPI shows
the scattering intensity of the impurities as function of momentum
transfer $\vec{q}$ between the initial and final wave vectors
\cite{Sprau2017}, one typically compares QPI patterns calculated from
theoretical band structures with the experimental one to reveal the nature of the underlying bands. Thus, QPI mainly focusses on the band dispersion.

We here take the opposite approach to gain complementary information.
In samples without defects, the Bloch states cause a LDOS that does
not vary from unit cell to unit cell (no QPI patterns). Instead, the
atomic part of the wave function causes LDOS variations within the
unit cell directly reflecting the quantum numbers $\lambda$. Thus, the
atomically resolved image of Fig. \ref{Fig1}a basically shows the
iso-surface of $\sum_\lambda |u_\lambda^{\vec{k}} (\vec{r})|^2$.  Figure
\ref{Fig1}c shows a topographic map of the size of 2$\times$2 unit
cells. The unit cell and its symmetry are indicated by solid back lines
(translational symmetry) and dashed lines (mirror planes). The white
dots represent the positions of the upper Se atoms. The image was
created by averaging over 16 individual unit cells using the
translational symmetry and by using the two mirror planes. This
procedure significantly reduces the statistical noise in the STM data. Besides the
Se atoms, a low intensity and finer structure becomes visible. Figure
\ref{Fig1}d displays an iso-LDOS map within the surface unit cell
calculated from first-principles. It was obtained by determining the surface of
constant LDOS in the vacuum in front of the surface corresponding to
the setpoint conditions of the STM experiment (see Methods). The
experimental and calculated maps agree reasonably well (see below for
more details).

Figure \ref{Fig2}a illustrates the energy dependence of the LDOS near
$\Delta$ recorded at 30 mK \cite{Balashov18}. When taking relatively
mild tunneling conditions for tip stabilization ($U$= 5\ mV and
$I$=320\ pA, green curve), the $dI/dV$ curves, which are proportional
to the LDOS, clearly shows a nodeless gap between $\pm\approx$150
$\mu$V, and two clear coherence peaks at $\approx$ 2.2 and 1.3 meV in
agreement with previous results on the anisotropic gap of FeSe
\cite{Sprau2017, Chen2017}. Thus, we can validate the full gap with our
energy resolution of $\approx$24 $\mu$eV. The two shoulders inside the
lowest energy coherence peaks stem from the anisotropic gap. Under
these mild tunneling conditions, we probe the evanescent Bloch
states relatively far from the topmost atoms. It is well established,
that far out the LDOS is dominated by s-electrons \cite{Hofer2003,Miyamachi2016} as s-electrons
decay the slowest into the vacuum. Note that due to the lifting of the continuous rotation symmetry by the crystal,
the angular momentum is not a strict quantum number and all orbitals
partially mix into s-states.  When, however, increasing the tunneling
conductance by approaching the tip further ($U$= 5\ mV and $I$=1\ nA,
blue curve), the contributions of orbitals with larger orbital
momentum to the LDOS will increase \cite{Miyamachi2016}. As a
consequence, the gap spectrum gains more structure. The shoulders
develop into coherence peaks and the wider coherence peaks split into
several peaks. As will be shown later, this is due to states of
different orbital character and different gaps. Most interestingly,
when recording $dI/dV$ spectra at low distance at different lateral
positions (see Fig. \ref{Fig2}b) the intensities of the features but
not their energy vary. Note that an increase of the intensity at
positive bias goes hand in hand with an increase at the same negative
bias. This hints for orbitally selective superconducting gaps.

In order to disentangle the orbital contributions, we carried out the
following experiments and calculations. First, we recorded $dI/dV$
spectra within 12 unit cells with a resolution of 11x11 points in the
individual unit cell. Similar to the topography, we use translational, rotational
and mirror symmetries to average the data. Figure \ref{Fig3}a shows
the averaged and symmetrized extracted LDOS. Note, however, that the
individual spectra vary within the unit cell, as will be discussed
below, but the peak positions observed in the low
current spectra can all be found in the averaged high-current spectrum.  Next, we
decompose the LDOS near $E_F$ from the DFT calculations into contributions of the different orbitals on the
afore-mentioned iso-plane of the total LDOS of Fig. \ref{Fig1}d. The
patterns within the unit cell are displayed in Figure \ref{Fig3}c. The calculations include
all s, p and d- orbitals. It can be seen, that the composition
of the LDOS varies dramatically within the unit cell for the different
orbitals. Note that due to the different
quantum numbers of the orbitals, no interference terms for the electrons
tunneling into the different orbitals (final states) is expected. Thus, the
LDOS can be written as a linear combination of the partial LDOS of the
different orbitals (see Supplementary). This gives the opportunity to
decompose the experimentally observed LDOS patters into their
constituents regarding the orbital quantum numbers.  Figure
\ref{Fig3}d displays pairs of the observed pattern and the simulated
pattern as well as the weights of the composition of the different orbitals
(bar graphs) at energies as indicated.  While the s-state still
dominates the current, clearly variations of the weights of the other
orbitals are observed. These are responsible for the variation of the
patterns. The experimental and simulated patterns agree well within
the capabilities of STM and our first-principles calculations. The
p-states are in general of higher intensity than the d-states.
Moreover, the d-states that extend more into the vacuum ($d_{z^2}$
more than $d_{xz}$ and $d_{yz}$) have a higher intensity. The in-plane
$d_{xy}$ and $d_{x^2-y^2}$ states have undetectable weight. This
agrees with the expectation of the spatial distribution of the orbitals. Figure \ref{Fig3}b plots the
contributions of all detected orbitals as function of energy. The
s-spectrum agrees nicely with the low-current spectrum of Fig.
\ref{Fig2}a. Most interestingly, the individual coherence peaks in the data of Fig. \ref{Fig3}b show
clear selectivity to specific orbitals. For example, the lowest
coherence peak at 0.28 meV is mostly of $p_x$ character, and the
highest peak at 1.08 meV is mostly of $p_z$ character and the shoulder
at 1.12 meV has a large $p_y$ character.  These states stem from the
p-orbitals of Se. Similarly, the contribution of the Fe d-states to the
different coherence peaks varies largely. The coherence peak at 0.82
meV is composed equally of $d_{xz}$ and $d_{yz}$. In the nematic phase, this is
characteristic for the hole pocket at the
$\Gamma$-point \cite{Watson_2015,Watson_20172, Kreisel_2015, Mukherjee_2015, Benfatto2018, Guterding2017, Yamakawa2016, Ishizuka2018}. Thus we can attribute this gap to
the hole pocket. The coherence peak at 1.12 meV shows a sizeable
contribution of $d_{yz}$ but not of $d_{xz}$. Thus, the symmetry is
broken which is characteristic for the elliptical electron pockets at
the M-point and we can associate the gap to the electron pocket.

We hope that our analysis regarding the p-orbitals of Se and their
contributions to the bands will stimulate further calculations on the
hybridization of the p- and d-states. This would allow to identify
more of the coherence peaks.
Finally, when the contribution of specific states peaks at a
certain bias voltage, QPI patterns recorded at that voltage may also
provide momentum information. As such, we see this approach as a
method to extend the use of STM in superconductivity research,
especially in combination with QPI.

\bibliographystyle{nature}
\bibliography{./FeSeuc}

\newpage

\noindent
{\bf Methods}

Single crystals of FeSe were prepared by chemical vapor transport
using elemental Fe and Se and a eutectic mix of the chlorine salts,
KCl and AlCl$_3$ in a constant temperature gradient \cite{Bohmer2016}.
On the single crystals, steel posts were glued and the samples were
transferred into ultra high vacuum (UHV). They were cleaved in UHV at
a 77 K and were directly inserted into the LT-STM followed by a rapid
cool down to the base temperature. STM tips were electrochemically
etched from thin W wires and were atomically cleaned by flashing in
UHV. $dI/dU$ spectra were obtained by numerical differentiation from
$I(U)$ curves to avoid modulation broadening. $dI/dU$ maps have been
recorded with a lock-in amplifier and a modulation voltage of
$U_\mathrm{rms}=28 \mu$V.  Experimental patterns in the unit cell were
fitted with calculated patterns using a least-square method.

First-principles calculations were performed using a self-consistent
Green function method within the multiple scattering theory specially
designed for semi-infinite systems such as surfaces and
interfaces~\cite{Luders2001}. The method utilizes the density
functional theory in the generalized gradient
approximation~\cite{Perdew1996}. The crystalline structure of the FeSe
bulk and surface was adopted from the literature. The vacuum
was modelled by layers with empty spheres. The potentials of 10 FeSe
units cells and 6 vacuum unit cells were calculated self-consistently
with proper boundary conditions for semi-infinite geometry. The LDOS was
calculated up to distances of 6{\AA} above the surface. Iso-LDOS surfaces were constructed from the
LDOS representing the experimental tunneling conditions using the Tersoff-Hamann approximation \cite{Tersoff1985}.

\noindent
{\bf Supplementary Information} is linked to the online version of the paper at www.nature.com/nature.

\noindent
{\bf Acknowledgements}
The authors acknowledge funding by the Deutsche Forschungsgemeinschaft
(DFG) under the grant WU 349/12-1, INST 121384/30-1 FUGG (W.W.), SCHM
1031/7-1 (J.S.) and SFB762/3 A04 (A.E.), as well as funding by the Alexander-von-Humboldt foundation (F.Y.).

\noindent
{\bf Author contribution}
F.Y., J.J. and W.W. conceived the experiments. F.Y. and J.J. carried
out the STM experiments. F.Y. and W.W. carried out data analysis.
S.F. and A.E. performed ab-initio calculations. T.W. and P.A. grew the
samples. J.S. and M.K. performed analytical calculations.  F.Y., J.J.
and W.W. wrote the draft. All authors contributed in discussions and
finalizing the manuscript.

\noindent
{\bf Author information}
Reprints and permissions information is available at
www.nature.com/reprints. The authors declare no competing financial
interests. Readers are welcome to comment on the online version of the
paper. Correspondence and requests for materials should be addressed
to F.Y. (francoyang1988@163.com).

\newpage

\begin{figure*}
  \centering
  \includegraphics[width=\columnwidth]{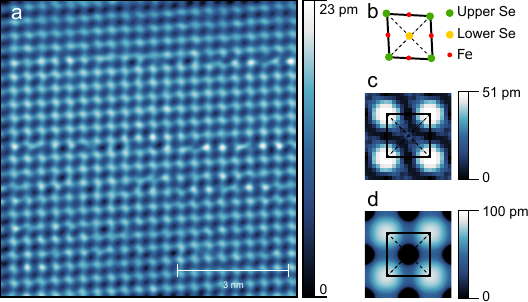}\\
  \caption{{\bf Structure of FeSe unit cell.} (a) STM-topography of
    impurity- and defect-free FeSe surface ($U$=15\ mV$\gg\Delta$,
    $I$=200\ pA). (b) Schematic drawing of the FeSe surface unit cell,
    which has $C_2$ rotational symmetry and two mirror planes (dashed
    lines). The nematic distortion of the unit cell is exaggerated in the figure. (c) STM-topography of FeSe of (2$\times$2) surface unit
    cells extension obtained by averaging from original data of larger
    range ($U$=5\ mV$>\Delta$, $I$=2\ nA). (d) Simulated STM
    topography obtained from the LDOS calculated from first-principles.}\label{Fig1}
\end{figure*}

\begin{figure*}
  \centering
  \includegraphics[width=\columnwidth]{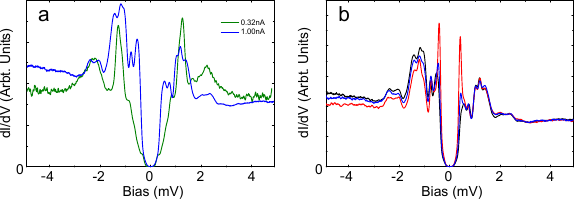}\\
  \caption{{\bf Differential conductance as function of tip-sample distance.} (a)Tunneling
    spectra recorded at $30$ mK at different tip-sample distances.
    Feedback conditions for the green curve: $U$=5\ mV, $I$= 0.32\ nA,
    blue curve: $U$=5\ mV, $I$=1.00\ nA. (b) Tunneling spectra at
    different lateral positions with feedback conditions $U$=5\ mV,
    $I$=1\ nA.}\label{Fig2}
\end{figure*}

\begin{figure*}
  \centering
  \includegraphics[width=13cm]{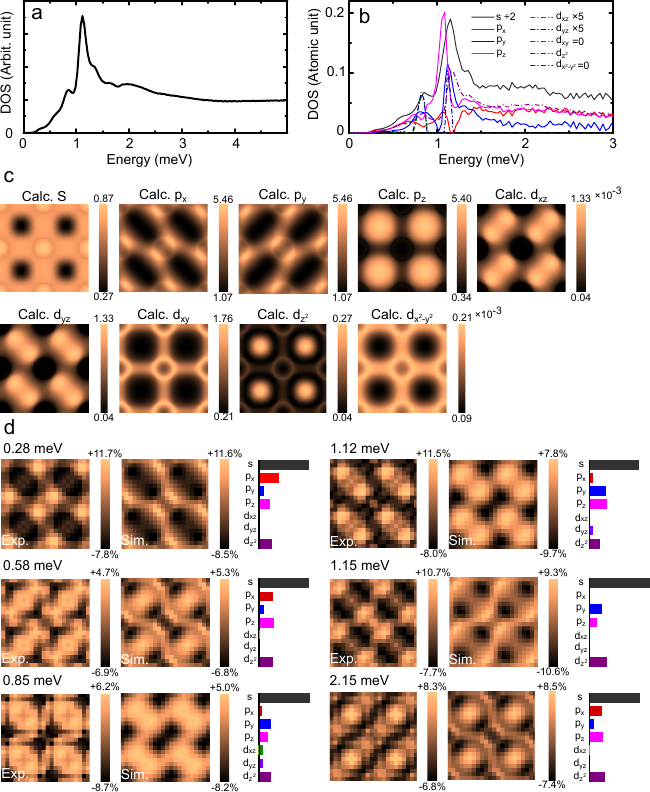}\\
  \caption{{\bf Orbitally resolved LDOS of the superconducting
      coherence peaks.}(a) Spatially averaged and symmetrized
    tunneling spectrum with feedback parameters $U$=5\ mV, $I$=2\ nA
    displaying several coherence peaks. (b) Fitted partial LDOS within
    the unit cell for s-, p- and d-states near the Fermi level
    revealing the orbital nature of the different peaks. (c) Partial
    LDOS of the different orbitals calculated on the iso-LDOS surface.
    This reflects the square of the Bloch wave function orbitally
    resolved. (d) Pairs of the experimental distribution of the LDOS
    in the unit cell in direct comparison with a fitted superposition
    of the calculated partial LDOS of c for energies as indicated. To
    the right of the pairs the fitting result of the relative
    contributions of the different orbitals is graphically displayed using the same colour code for the orbitals as in b.
    The experimental and theoretical distributions in the unit cell
    match nicely. From these fits, the partial LDOS of b was
    constructed. }\label{Fig3}
\end{figure*}

\clearpage
\title{Orbitally resolved superconductivity in real space: FeSe}


\author{Fang Yang}
%\email{francoyang1988@163.com}
\affiliation{Institute for Nanoelectronic Devices and Quantum Computing, Fudan University, Songhu Rd. 2005, 200438 Shanghai, P.R. China}
\affiliation{Physikalisches Institut, Karlsruhe Institute of Technology, Wolfgang-Gaede Str. 1, 76131 Karlsruhe, Germany}

\author{Jasmin Jandke}
\affiliation{Physikalisches Institut, Karlsruhe Institute of Technology, Wolfgang-Gaede Str. 1, 76131 Karlsruhe, Germany}

\author{Peter Adelmann}
\affiliation{Institut f\"ur Festk\"orperphysik, Karlsruhe Institute of Technology, 76344 Karlsruhe, Germany}

\author{Markus J. Klug}
\affiliation{Institut f\"ur Theorie der Kondensierten Materie, Karlsruhe Institute of Technology, Wolfgang-Gaede Str. 1, 76131 Karlsruhe, Germany}

\author{Thomas Wolf}
\affiliation{Institut f\"ur Festk\"orperphysik, Karlsruhe Institute of Technology, 76344 Karlsruhe, Germany}

\author{Sergey Faleev}
\affiliation{Max-Planck-Institut f\"{u}r Mikrostrukturphysik, Weinberg 2, D-06120 Halle, Germany}

\author{J\"{o}rg Schmalian}
\affiliation{Institut f\"ur Theorie der Kondensierten Materie, Karlsruhe Institute of Technology, Wolfgang-Gaede Str. 1, 76131 Karlsruhe, Germany}
\affiliation{Institut f\"ur Festk\"orperphysik, Karlsruhe Institute of Technology,
76344 Karlsruhe, Germany}
\author{Matthieu Le Tacon}
\affiliation{Institut f\"ur Festk\"orperphysik, Karlsruhe Institute of Technology,
76344 Karlsruhe, Germany}

\author{Arthur Ernst}
\affiliation{Max-Planck-Institut f\"{u}r Mikrostrukturphysik, Weinberg 2, D-06120 Halle, Germany}
\affiliation{Institute for Theoretical Physics, Johannes Keppler University Linz, Altenberger Stra{\ss}e 69, 4040 Linz, Austria}

\author{Wulf Wulfhekel}
\affiliation{Physikalisches Institut, Karlsruhe Institute of Technology, Wolfgang-Gaede Str. 1, 76131 Karlsruhe, Germany}
%\affiliation{State Key Laboratory of Surface Physics and Department of Physics, Fudan University, Shanghai 200433, China}

\date{\today}

\maketitle

\makeatletter
%%%%%%%%%%%%%%%%%%%%%%%%%%%%%% Textclass specific LaTeX commands.
\newcommand{\lyxrightaddress}[1]{
	\par {\raggedleft \begin{tabular}{l}\ignorespaces
	#1
	\end{tabular}
	\vspace{1.4em}
	\par}
}

\makeatother

\subsection*{Vanishing of orbital interference terms in the local density of states }

In the following, we provide evidence for the assumption that no interference
terms for tunneling electrons tunneling into orbitals exist. It implies
that the local density of states is given by the sum of partial density
of states of different orbitals $\alpha$,

\begin{equation}
\rho\left(\mathbf{r},\omega\right)=\sum_{\alpha}\rho_{\alpha}\left(\mathbf{r},\omega\right).\label{eq:localDOSres-1}
\end{equation}
Our analysis assumes tetragonal crystal systems and is therefore applicable
to bulk or surface electron systems hosted in FeSe as discussed in
the main text. The result
is furthermore based on the assumption that there is no (weak) spin
orbit coupling. Finite spin orbit coupling may potentially alter the
result such that interference terms are present. This aspect will
be discussed in more detail at the end of this section.

The following analysis relies on group theoretical arguments and is
therefore independent on the exact microscopic model. We start by
expressing the local density of states probed in STM measurements
in terms of the retarded single-particle correlator
\begin{align}
\rho\left(\mathbf{r},\omega\right) & =-\frac{1}{V}\frac{1}{\pi}\sum_{\sigma}\text{Im}G_{\sigma\sigma}^{R}\left(\mathbf{r},\mathbf{r},\omega\right).\label{eq:localDOS}
\end{align}
with spin index $\sigma$. Here, averaging over spin degrees of freedom
assumes implicitly spin-rotational symmetry of the tip/probe system.
Expressed in localized Wannier orbitals $\phi_{\alpha}\left(\mathbf{r}\right)$
forming a complete set of single-particle states, the single-particle
propagator reads
\begin{equation}
G_{\sigma\sigma'}^{R}\left(\mathbf{r},\mathbf{r}',\omega\right)=\sum_{\alpha\beta}\sum_{ij}G_{\sigma\sigma',\alpha\beta}^{R}\left(\mathbf{R}_{i},\mathbf{R}_{j},\omega\right)\phi_{\alpha}\left(\mathbf{r}-\mathbf{R}_{i}\right)\phi_{\beta}^{*}\left(\mathbf{r}'-\mathbf{R}_{j}\right)\label{eq:Gr}
\end{equation}
with orbital indices $\alpha,\beta$ and the location of Bravais lattice
site $\mathbf{R}_{i}$ (the extension to a $n$-dimensional basis
is straightforward). In the following, we assume that the Wannier
orbitals decay exponentially on length scales of the lattice spacing.
It is therefore sufficient to consider the overlap of wave functions
at the same lattice site $i=j$ only. In what follows, we will show
that orbital off-diagonal terms, $\alpha\neq\beta$, of the single-particle
correlator vanish,
\begin{equation}
G_{\sigma\sigma',\alpha\neq\beta}^{R}\left(\mathbf{R}_{i},\mathbf{R}_{i},\omega\right)=0,\label{eq:cond}
\end{equation}
which leads to the vanishing of orbital interference terms in Eq.
(\ref{eq:localDOSres-1}).

We consider the single-particle Hamilton operator represented in the
Bloch basis by
\begin{equation}
H=\sum_{\mathbf{k}}\sum_{\alpha\beta}\epsilon_{\alpha\beta}\left(\mathbf{k}\right)d_{\alpha,\mathbf{k}}^{\dagger}d_{\beta,\mathbf{k}},
\end{equation}
with electronic creation-/annihilation operator $d_{\alpha,\mathbf{k}}^{\left(\dagger\right)}$
and orbital dependent dispersion relations $\epsilon_{\alpha\beta}\left(\mathbf{k}\right)$.
Electronic single-particle states are labeled by the crystal momentum
$\mathbf{k}$ and orbital indices $\alpha,\beta$. We rewrite the
Hamilton operator using operator bilinear forms $T^{\left(\Gamma\right)}\left(\mathbf{k}\right)=\sum_{\alpha\beta}d_{\alpha,\mathbf{k}}^{\dagger}\lambda_{\alpha\beta}^{\left(\Gamma\right)}d_{\beta,\mathbf{k}}$
with $(\lambda_{\alpha\beta}^{\left(\Gamma\right)})$ being matrices
in orbital space, which transform under the point symmetry operations
of the underlying lattice according to the one-dimensional irreducible
representation $\Gamma\in\left\{ A_{1},A_{2},B_{1},B_{2}\right\} $,
\begin{equation}
H=\sum_{\mathbf{k}}\sum_{\Gamma}h^{\left(\Gamma\right)}\left(\mathbf{k}\right)T^{\left(\Gamma\right)}\left(\mathbf{k}\right).
\end{equation}
Dispersion relations are contained in $h^{\left(\Gamma\right)}\left(\mathbf{k}\right)$
which have well defined transformation behaviors depending on $\Gamma$.
The single-particle correlator is consequently given by

\begin{equation}
G_{\alpha\beta}\left(z,\mathbf{k}\right)=\big[z-\sum_{\Gamma}h^{\left(\Gamma\right)}\left(\mathbf{k}\right)\lambda_{\alpha\beta}^{\left(\Gamma\right)}\big]^{-1}.
\end{equation}
We now distinguish between contributions of trivial ($\Gamma=A_{1}$)
and non-trivial ($\Gamma\neq A_{1}$) transformation behavior and
rewrite the the previous equation by introducing the orbital matrix
$\hat{G}=(G_{\alpha\beta})$ as
\begin{align}
\hat{G}\left(z,\mathbf{k}\right) & =\hat{G}_{0}\left(z,\mathbf{k}\right)+\hat{G}_{0}\left(z,\mathbf{k}\right)\hat{V}\left(\mathbf{k}\right)\hat{G}\left(z,\mathbf{k}\right)\label{eq:}\\
 & =\hat{G}_{0}\left(z,\mathbf{k}\right)+\hat{G}_{0}\left(z,\mathbf{k}\right)\hat{V}\left(\mathbf{k}\right)\hat{G}_{0}\left(z,\mathbf{k}\right)+\dots\label{eq:Grow}
\end{align}
with $\hat{G}_{0}\left(z,\mathbf{k}\right)=\left[z-h^{(A_{1})}\left(\mathbf{k}\right)\lambda^{(A_{1})}\right]^{-1}$
and $\hat{V}=\sum_{\Gamma\neq A_{1}}h^{\left(\Gamma\right)}\left(\mathbf{k}\right)\lambda^{\left(\Gamma\right)}$.

We now consider the sum over crystal momenta of the correlator, $\sum_{\mathbf{k}}G_{\alpha\beta}(\mathbf{k},z)=G_{\alpha\beta}(\mathbf{R}_{i},\mathbf{R}_{i},z)$,
which is identical to its previously introduced real space version.
By inspecting Eq. (\ref{eq:Grow}), we find that $\sum_{\mathbf{k}}G_{\alpha\beta}(z,\mathbf{k})=0$
for $\alpha\neq\beta$, where each term vanishes individually. This
finding is traced back to the fact that either (I) the resulting matrix
in orbital space is diagonal, or (II) the sum over crystal momenta
of an non-trivially transforming function vanishes, $\sum_{\mathbf{k}}G^{\left(\Gamma\right)}\left(\mathbf{k}\right)\big|_{\Gamma\neq A_{1}}=0$.
In particular, the first term vanishes because of (I); the second
because of (II), $\sum_{\mathbf{k}}[h^{(A_{1})}(\mathbf{k})]^{2}h^{(\Gamma\neq A_{1})}\left(\mathbf{k}\right)=0$;
the third because of (I) for $\Gamma=\Gamma'$ and because of (II)
for $\Gamma\neq\Gamma'$, $\sum_{\gamma\mathbf{k}}h^{(\Gamma)}\left(\mathbf{k}\right)h^{(\Gamma')}\left(\mathbf{k}\right)\lambda_{\alpha\gamma}^{(\Gamma)}\lambda_{\gamma\beta}^{(\Gamma')}\big|_{\Gamma\neq\Gamma'}=0$.
The identical reasoning can be applied to any higher order term by
noticing that if the orbital matrix is non-diagonal the product of
$h$'s contains an odd number of non-trivially transforming functions.
This eventually implies Eq. (\ref{eq:cond}).

In the case of finite spin orbit coupling, when the spin degrees of
freedom cease to be good quantum numbers, the result is potentially
altered. In the presence of inversion symmetry, which holds true for
electronic bulk states in the tetragonal crystal system, all bands
are still doubly degenerate and the Hamilton operator can be rewritten
in terms of bilinear form transforming according to one-dimensional
representations. However, this potentially changes when the inversion symmetry
is broken, which is especially the case when electronic surface states being probed
in STM measurements. Here, the surface state band's degeneracy may be lifted. Furthermore,
the orbital character of electronic states of the STM tip has to be
taken into account. The average over spin states in Eq. (\ref{eq:localDOS})
is replace by a weighted average depending on the STM tip's orbital
character and a cancelation of contributions is not guaranteed. For
this more complex scenario, the previously drawn conclusion manifesting
in Eq. (\ref{eq:localDOSres-1}) loses its rigidity and orbital interference
terms may appear as function of the strength of spin-orbit interaction.

\end{document}